\documentclass{aa}
\usepackage{txfonts}
\usepackage{lscape,graphicx}
\usepackage{natbib}
\bibpunct{(}{)}{;}{a}{}{ }

\begin{document}

\title{RX\thinspace J0042.3+4115: a stellar mass black hole\\ binary identified in M31}
\author{ R. Barnard\inst{1}
    \and J. P. Osborne\inst{2}
    \and U. Kolb\inst{1}
     \and K. N. Borozdin\inst{3}}

\offprints{R. Barnard, \email{r.barnard@open.ac.uk}}

\institute{ The Department of Physics and Astronomy, The Open University, Walton Hall, Milton Keynes, MK7 6BT, U.K.
     \and The Department of Physics and Astronomy, The University of Leicester, Leicester, LE1 7RH, U.K.
     \and Los Alamos National Laboratory, P.O. Box 1663, Mail Stop D436, NIS-2, Los Alamos, MN 87545}
\date{ Received 6 February  2003 / Accepted 17 April 2003}

\abstract{

Four XMM-Newton X-ray observations of the central region of the \object{Andromeda Galaxy} \object{(M31)} have revealed an X-ray source that varies in luminosity over $\sim$1--3\, 10$^{38}$ erg s$^{-1}$ between observations and also displays significant variability over time-scales of a few hundred seconds. The power density spectra of lightcurves obtained in the 0.3--10 keV energy band from the three EPIC instruments on board XMM-Newton are typical of disc-accreting X-ray binaries at low accretion rates, observed in neutron star binaries only  at much lower luminosities ($\sim 10^{36}$ erg s$^{-1}$). However X-ray binaries with  massive black hole primaries have  exhibited such power spectra for luminosities $>$10$^{38}$ erg s$^{-1}$.  We discuss alternative possibilities  where  RX\thinspace J0042.3+4115 may be a background AGN or foreground object in the field of view, but conclude that it is located within M31 and hence use the observed power spectra and X-ray luminosities to identify the primary as a black hole candidate.

 }

\titlerunning{ Black hole binary identified  in M31}
\maketitle

\section{Introduction}

 At 760 kpc \citep{vdb00} the \object{Andromeda Galaxy (M31)} is the nearest spiral galaxy to our own. The X-ray emission from the direction of M31 is dominated by point sources, mostly X-ray binaries (XB), with a mixture of supernova remnants, foreground stars and background AGN. 
The intensity variations on short and long time-scales in the X-ray sources in \object{M31} provide vital clues to their nature, and in some cases, in combination with X-ray spectral information, they can be classified from the X-ray observations alone.
 The study of variability in X-ray sources in external galaxies has been limited by the sensitivity of previous observatories, but the three large X-ray telescopes  on board XMM-Newton make such a project possible down to a luminosity of $\sim10^{36}$ erg s$^{-1}$; the combined effective area of these telescopes is the largest of any X-ray telescope imaging above 2 keV. A recent Chandra survey has revealed that $\sim$ 50\% of the X-ray sources in M31 vary over long time-scales \citep{K02}, several transients have been discovered in XMM-Newton observations \citep[e.g. ][]{tbp01,tpb02,osb01} and  periodic dipping was observed with XMM-Newton in the M31 globular cluster source \object{Bo158} by \citet{tru02}.  

 The J2000 X-ray position of \object{RX J0042.3+4115}, named following  \citet{S97} but first identified by \citet{tf91},  was determined to be 00$^{\rm h}$~42$^{\rm m}$~22$\fs$919 +41$\degr$ 15$\arcmin$ 35$\farcs$14 in a  recent Chandra survey of M31 \citep{K02}. It was not associated with known supernova remnants, foreground objects, globular clusters or background galaxies \citep{K02}.  RX J0042.3+4115 shows luminosities in the 0.3--10 keV band of a few 10$^{38}$ erg s$^{-1}$, and significant variations in colour and intensity over time-scales of a few hundred seconds (Sect.~\ref{res}); in our galaxy, such behaviour is exhibited in only a handful of XB, all of which are bright microquasars with neutron star primaries  \citep[Z-sources, e.g.][]{hv89,fen02} or transient microquasars with suspected black hole primaries \citep[e.g.][]{bel00,fen02}.

In this paper we propose that  RX J0042.3+4115 is a black hole binary. The case is summarised below and expounded in the subsequent sections.
The power density spectra (PDS) of the $\sim$200 Galactic disc-accreting  X-ray binaries (XB)  are determined by their accretion rates, and hence luminosities; nearly  all disc accreting XB are low mass X-ray binaries (LMXB), but some high mass X-ray binaries (e.g. \object{Cygnus\, X-1}) also exhibit disc accretion. The PDS of low accretion rate, disc accreting XB  are approximately flat at frequencies  $<$0.01--1 Hz, then break to a steep power law; these characteristics are present irrespective of the primary \citep{wv99}. At high accretion rates the PDS are varied but are  distinct from the low accretion rate PDS in that the low frequency noise  is described by a power law without the break \citep{hv89}. In Galactic neutron star LMXB,  low accretion rate PDS are seen in sources with luminosities of a few 10$^{36}$ erg s$^{-1}$; in contrast we find that RX J0042.3+4115 exhibits a low accretion rate PDS at luminosities of $\sim$1--3 10$^{38}$ erg s$^{-1}$. This is close to the  Eddington luminosity of an accreting neutron star [$\sim$2 10$^{38}$ erg s$^{-1}$, \citep{fkr92}]. If RX\thinspace J0042.3+4115 was a neutron star binary, a high accretion rate would be required for the observed luminosity and the characteristic low accretion rate PDS would not be seen. Hence, the primary in RX\thinspace J0042.3+4115 is unlikely to be a neutron star and so must be a black hole.

The next section reviews the properties of  Galactic black hole binaries to provide a context for the subsequent discussion. A summary of observations and analysis techniques follows, then the luminosities and power spectral breaks observed in RX\thinspace J0042.3+4115 are presented in the Results section. The case for a black hole primary in RX\thinspace J0042.3+4115 is presented in the final section.

\section{Galactic black hole binaries}

Black hole binaries can be classified as persistent or transient, and each can exist in any one of three states- the low/hard state, the high/soft state and intermediate/very high state. The X-ray spectrum in the high/soft state is black-body dominated, while the low/hard spectrum is dominated by a power law with a spectral index $\sim$1.5. The intermediate/high state exhibits strong black body and power law features. As suggested by their names, these states were originally supposed to be related to the mass accretion rate and hence luminosity; however, the same ``state'' can be observed at extremely different flux levels \citep[e.g.][]{hom01}.
The four Galactic persistent black hole binaries (\object{Cygnus\thinspace X-1}, \object{GX\thinspace 399$-$4}, \object{1E\thinspace 1740.7$-$2942} and \object{GRS\thinspace 1758$-$258}) spend most of their time in their low/hard state \citep[][ and references within]{fen02}. Meanwhile, transient black hole binaries can switch states very rapidly; \citet{bel00} have identified 12 separate classes of behaviour in the microquasar \object{GRS\thinspace 1915+105} which involve transitions between two or all three of the states. 

Microquasars are compact accreting binary systems with low mass secondary stars that exhibit relativistic outflows, or jets; the similarity with quasars gives them their name \citep{fen02}. Relativistic jets from binaries were first discovered in \object{SS\, 433} \citep{spen79,hj81a,hj81b}, and have since been discovered in $\sim$30 Galactic XB \citep{fen02}; 11 microquasars are neutron star binaries and 19 are black hole binaries \citep{fen02}. As yet, the only Galactic black hole binaries to exhibit luminosities $>$10$^{38}$ erg s$^{-1}$ have been microquasars. Hence, they may be the closest Galactic analogue to RX\thinspace J0042.3+4115.

\begin{figure}[!t]
\resizebox{\hsize}{!}{\includegraphics{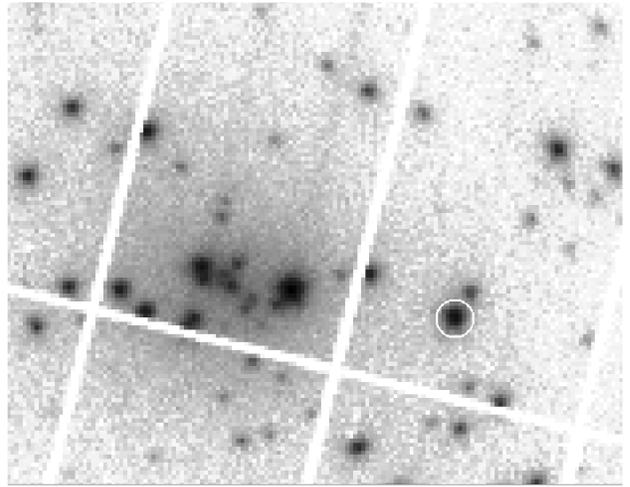}}
\caption{ A detail of the 0.3--10 keV PN image from the June 2001  XMM-Newton observation of the core of M31; north is up, east is left.  The image is 11$\arcmin$ across, its intensity is log scaled.  RX\thinspace J0042.3.9+4115 is circled in white, the circle defining the source extraction region.}\label{obx3}
\end{figure}

\begin{figure*}[!t]
\centering
\resizebox{\hsize}{!}{\includegraphics{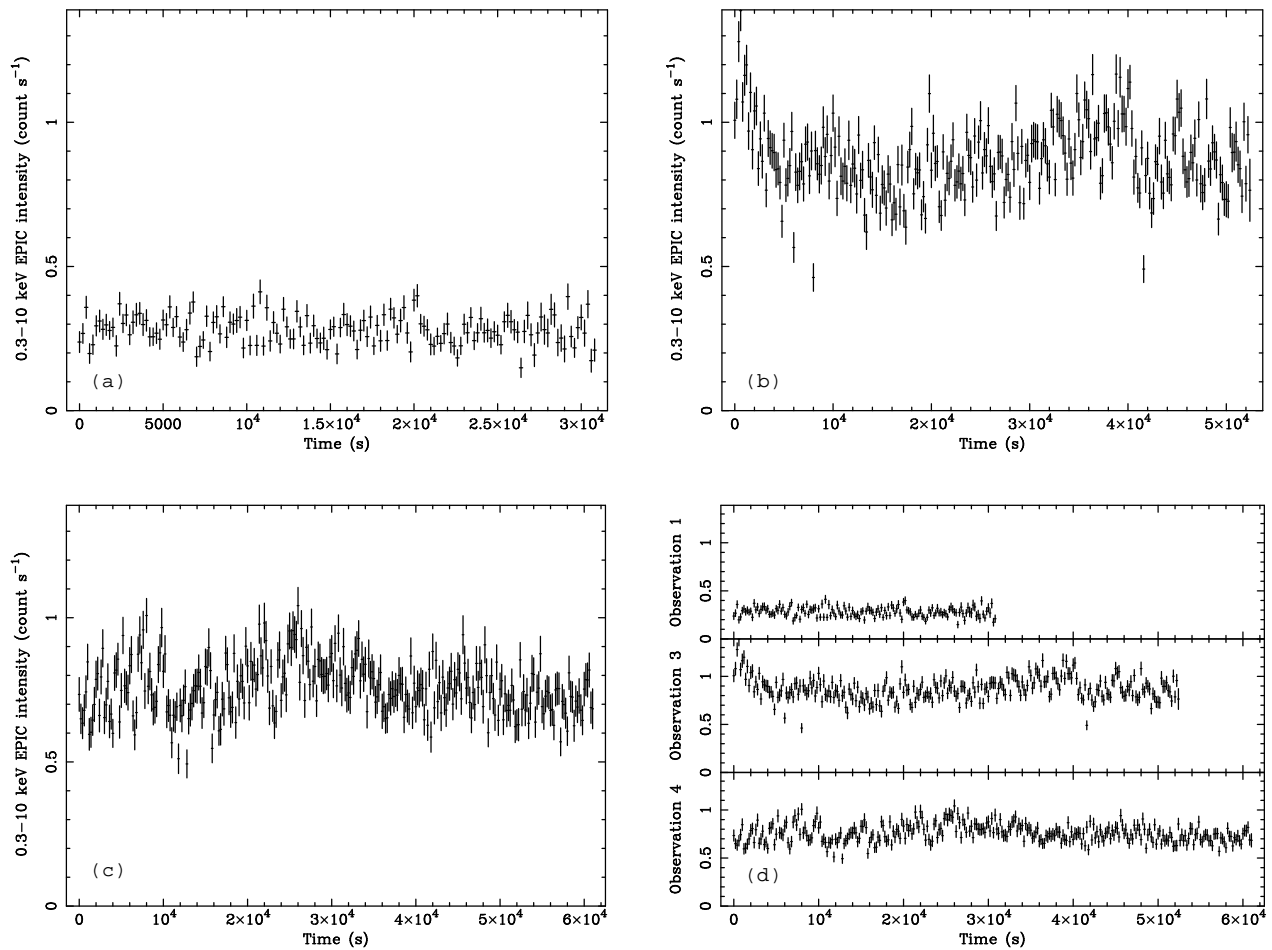}}
\caption{ Combined EPIC lightcurves of RX J0042.3+4115 in the 0.3--10 keV  energy band  from observations 1 ({\bf a}), 3 ({\bf b}) and 4 ({\bf c}); the lightcurves have 200 s binning. The $\chi^{2}$/dof for the best fit lines of constant intensity for observations 1, 3 and 4 are 287/154, 329/125 and 798/306 respectively. Although each lightcurve is highly variable, there appears to be no pattern to their behaviour.  The three lightcurves are presented with the same time and intensity scales in (d) for easier comparison.}\label{13_1}
\end{figure*}

\section{The Observations}

Four observations were made of the core of M31 with XMM-Newton; data were analysed from the three imaging  instruments, one PN \citep{stru01} and two MOS \citep{turn01}, which share a common circular field-of view with a 30$\arcmin$ diameter \citep{jan01}. The MOS detectors have a usable energy range of 0.3--10 keV, a point spread function FWHM of 5$\arcsec$, an energy resolution of $\sim$70 eV and a time resolution of 2.6 s in the ``full frame'' mode used for these observations \citep{turn01}. The PN detector has broadly similar characteristics, but with approximately twice the effective area of an individual MOS. Full frame mode was used, giving a  time resolution of 0.0734 s \citep{stru01}. The combined effective area is $\sim$2400 cm$^{2}$ at 1keV \citep{jan01}.
A journal of the  XMM-Newton observations is given in Table~\ref{journ}, which details the date, duration and filter used for each observation.  

\begin{table}[!t]
\centering
\caption{Journal of XMM-Newton observations of the \object{M31} core}\label{journ}
\begin{tabular}{lllll}
\noalign{\smallskip}
\hline
\noalign{\smallskip}

Observation & Date & Duration & Filter\\
\noalign{\smallskip}
\hline
\noalign{\smallskip}
1 (rev 0100) &  25 July 2000& 34 ks& Medium \\
2 (rev 0193)& 27 December 2000 & 13 ks&  Medium\\
3 (rev 0285)& 29 June 2001& 56 ks &Medium \\
4 (rev 0381)& 6 January 02 & 61 ks&  Thin\\
\noalign{\smallskip}
\hline
\noalign{\smallskip}

\end{tabular}
\end{table}

\section{Analysis}

Lightcurves were obtained in the 0.3--2.5, 2.5--10 and 0.3--10 keV energy bands, at 0.0734 s resolution from the PN and 2.6 s resolution from the MOS cameras. The  data were filtered to select only good events (FLAG = 0) with {\sc sas} version 5.3.3\footnote{http://heasarc.gsfc.nasa.gov/docs/xmm/uhb/}
 and analysed with the {\sc heasoft}  utility suite\footnote{ http://heasarc.gsfc.nasa.gov/docs/software/lheasoft/}; only events in the energy band 0.3--10 keV with PATTERN $<$ 4 or 12, for PN and MOS respectively, were selected. Background flaring was present in observations 1, 3 and 4, and so intervals in which the total PN count rate exceeded 60 count s$^{-1}$ or the count rate of a single MOS detector exceeded 30 count s$^{-1}$ were rejected; only observation 3 exceeded these thresholds, reducing the useful exposure to 27 ks.  There are $\sim$120 X-ray point sources in the field of view, and the  extraction regions for our source was limited to 20$\arcsec$ to avoid contamination by a neighbouring point source; this encircles $\sim$70\% of the flux. The PN image of the Core of M31 from the June 2001  XMM-Newton observation is shown in Fig.~\ref{obx3} with RX\thinspace J0042.3+4115 circled. 

For each lightcurve on source, a source-free background lightcurve was extracted from a nearby region on the same CCD and at a similar off-axis angle,   the same filtering criteria were used. The fractional exposure of the source was estimated from the exposure map, and  the lightcurves were corrected for the overall effects of deadtime and vignetting using {\sc fcalc}.

Additionally, PN source and background spectra were obtained from all good-time data for observations 1, 3 and 4; in observation 2, the source fell in a chip gap.   Response matrices and ancillary response files were generated for each spectrum, with {\sc rmfgen} and {\sc arfgen} respectively. The spectra were grouped to have a minimum of 50 counts per bin, data below 0.3 keV or greater than 10 keV were ignored in the subsequent analysis.

\section{Results}
\label{res}

The combined, exposure corrected, background subtracted  0.3 -- 10 keV EPIC lightcurves of RX\thinspace J0042.3+4115 from observations 1, 3 and 4 are presented in Fig.~\ref{13_1}; these lightcurves are clearly variable, but display no regular behaviour. The lightcurves vary by $\sim$30\% within each observation, while  the mean intensity of the lightcurve from observation 1 is a factor of $\sim3$ lower than the mean intensities in observations 3 and 4. 
\begin{table*}[!t]
\caption{ Best fit parameters for models used to fit the PN 0.3--10 keV spectra of RX\thinspace J0042.3+4115 in observations 1, 3 and 4; the letters (A, B, C) refer to the model used (as described in the text) and the numbers refer to the observation number. N$_{\rm H}$ is the equivalent line-of-sight column density of hydrogen (in cm$^{-2}$); kT is the blackbody temperature expressed in keV; n$_{\rm B}$ is the normalisation of the blackbody component: 10$^{39}$  erg s$^{-1}$ at 10 kpc for  blackbody, R$_{\rm in}^{2}$ cos $\theta$ for a disc blackbody of inner disc radius R$_{\rm in}$ observed at an angle $\theta$; $\Gamma$ is the spectral index of the power law component; n$_{\rm P}$ is the normalisation of the power law, in terms of photon cm$^{-2}$ keV$^{-1}$ s$^{-1}$ at 1 keV. F$^{0.3-10}$ and  L$^{0.3-10}_{760 \rm kpc}$ are the unabsorbed flux and luminosity at 760 kpc respectively in the 0.3 -- 10 keV energy band.}\label{specres}
\centering
\begin{tabular}{llllllllllll}
\noalign{\smallskip}
\hline
\noalign{\smallskip}
{\bf Model} & {\bf N$_{\rm H}$}& {\bf kT }& {\bf n$_{\rm B}$} & {\bf $\Gamma$}& {\bf n$_{P}$}  & {\bf $\chi^{2}$/d.o.f}&  {\bf F$^{0.3-10}$}& {\bf L$^{0.3-10}_{760 \rm kpc}$}\\
 & /10$^{22}$ &  &   & &   &  & /10$^{-12}$  & /10$^{37}$ \\
 \noalign{\smallskip}
\hline
\noalign{\smallskip}
A1 & 0.18 & \dots & \dots & 1.66 & 1.5 10$^{-4}$ & 104/91 & 1.08 & 7.5\\
B1 & 0.25 & 1.69 & 4.3 10$^{-6}$ & 2.2 & 1.75 10$^{-4}$ & 102/89 & 1.22 & 8.5 \\
C1 & 0.23 & 3.33 & 2.4 10$^{-4}$ & 2.2 & 1.39 10$^{-4}$ & 102/89 & 1.2 & 8.2\\
A3 & 0.16 & \dots & \dots & 1.79 & 4.7 10$^{-4}$ & 181/172 & 3.0 & 21\\
B3 & 0.25 & 1.32 & 1.36 10$^{-5}$ & 2.55 & 5.65 10$^{-4}$  & 162/170 & 3.8 & 24 \\
C3 & 0.31 & 2.28 & 3.59 10$^{-3}$ & 3.32 & 5.25 10$^{-4}$ & 162/170 & 4.9 & 34\\
A4 & 0.18 & \dots & \dots &1.82 & 4.9 10$^{-4}$&  352/190 & 3.0 & 21 \\
B4 & 0.23 & 1.23 & 1.16 10$^{-4}$ & 2.43 & 5.09 10$^{-4}$ & 229/188 & 3.4 & 24\\C4 & 0.30 & 2.03 & 5.79 10$^{-3}$& 3.4 & 4.47 10$^{-4}$ & 231/188 & 4.7& 33\\
 \noalign{\smallskip}
\hline
\noalign{\smallskip}
\end{tabular}
\end{table*}
Estimates of the luminosity of RX\thinspace J0042.3+4115 in each observation were obtained by fitting the background-subtracted PN spectra with three simple models. Model A is a simple power law, model B is  blackbody + power law and model C is a disc blackbody + power law; all models include the effects of photo-electric absorption from material in the line of sight. Although the parameters of the models have limited physical  significance, since the spectra are not of high enough quality to discriminate between models,  a reasonable  fit to a spectrum will provide a good estimate to the 0.3--10 keV flux and  luminosity. The best fit results are presented in Table~\ref{specres}. We obtain 0.3--10 keV luminosities of $\sim$8, $\sim$30 and $\sim$30 $\times 10^{37}$ erg s$^{-1}$ in observations 1, 3  and 4 respectively.  We see that Model A cannot fit the spectrum from observation 4, and thus must be be rejected for all observations. The average Galactic line of sight absorption in the direction of  M31 is 7 10$^{20}$  atom cm$^{-2}$ \citep{stark92} while the modelled absorption of RX\thinspace J00422.3+4115 is 20--30 10$^{20}$ atom cm$^{-2}$, as seen in X-ray sources in M31 \citep{K02}. This suggests that 
RX\thinspace J00422.3+4115 is indeed in M31, or else has intrinsic absorption resembling the absorption found in M31 by coincidence.

\begin{figure}[!t]
\centering
\resizebox{\hsize}{!}{\includegraphics[angle=270]{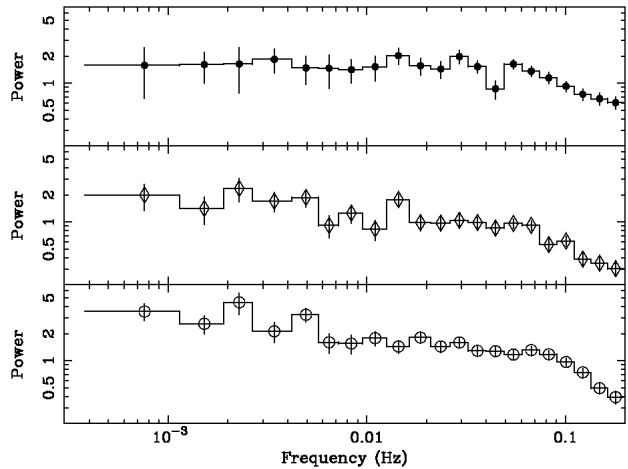}}
\caption{Power density spectra of RX\thinspace J0042.3+4115 from observations 1, 3 and 4 are displayed in the top, central and bottom panels respectively. The PDS of observation 1 is clearly flat below 0.04 Hz and a steep power law above 0.06 Hz, consistent with the PDS of low accretion rate XB \citep{wv99}; indeed, a break is seen in the PDS of all three observations, below which the PDS is approximately flat.} \label{pds}
\end{figure}

Leahy normalised \citep{lea83} power density spectra (PDS)  with (5.2 s)$^{-1}$  resolution were obtained from the 0.3 -- 10 keV EPIC  lightcurves from observations 1, 3 and 4 using {\sc powspec};  the PDS were averaged over multiple spectra of 512 bins  and the expected noise was subtracted; the PDS data were geometrically binned. The three power spectra are presented in Fig~\ref{pds}. We find that the PDS from all three observations are $\sim$ flat at low frequencies, and break to a steep power law at frequencies of $\sim$0.06--0.1 Hz. To establish the significance of the breaks, each PDS was modelled with both a simple power law  and  a broken power law. The parameters for the broken power law model were the cut-off frequency, $\nu_{\rm c}$, the power at that frequency, the index for frequencies lower than $\nu_{\rm c}$, $\alpha_1$, and the index for frequencies higher than $\nu_{\rm c}$, $\alpha_2$. Table~\ref{pdffit}  lists the parameters and goodness of fit of the two models for each PDS; clearly the simple power law model cannot acceptably fit any of the PDS, and all PDS are well fitted by the broken power law model, hence the PDS really do resemble those of low accretion rate binaries.

\begin{center}
\begin{table}[!b]
\centering
\caption{Fits of simple power law models (PL$x$) and broken power law models (BPL$x$) for observation $x$ (1, 3 and 4); $\alpha_1$ and $\alpha_2$ are the spectral indices before and after the break frequency ($\nu_{\rm c}$). Numbers in parentheses are the uncertainties in the last significant figure.}\label{pdffit}
\begin{tabular}{lllll}
\noalign{\smallskip}
\hline
\noalign{\smallskip}
Model & $\alpha_1$ & $\nu_{\rm c}$ (Hz) & $\alpha_2$ & $\chi^{2}$/dof\\
\noalign{\smallskip}
\hline
\noalign{\smallskip}
PL1 & $-0.11$(1) & \dots &\dots & 21.8/11\\
BPL1 & $-0.04$(2) & 0.05(1)& $-0.21(6)$ & 5.8/9\\
PL3 & $-0.35(2)$ & \dots & \dots & 62.2/11\\
BPL3 & $-0.24$(3) & 0.09(1)& $-$1.5(4)& 8.6/9\\
PL4 & $-$0.35(3) & \dots &\dots & 45.7/11\\
BPL4 & $-$0.27(4) & 0.05(3) & $-$0.5(2) & 13.7/9\\
\noalign{\smallskip}
\hline
\noalign{\smallskip}
\end{tabular}

\end{table}
\end{center}

To establish whether the break is an artifact of the instrumentation, or inherent to RX\thinspace J0042.3+4115, the PDS of seven  bright X-ray sources identified with globular clusters by \citet{K02} were examined; as they are likely to be high accretion rate neutron star binaries their PDS should not contain a break in the observed frequency range. Although here is no reason to expect   instrumental modulation of the true PDS  that would produce the breaks we have observed, these checks were nevertheless made to be sure.  For each globular cluster source, combined  EPIC lightcurves from observation 4 were made and a PDS was obtained. Luminosities of the globular cluster sources were estimated from fits to PN spectra of  simple power law models; the spectral channels were binned for a minimum of 20 counts per bin. Table~\ref{gc} lists the best fit model parameters for the globular cluster sources.  The luminosities of the globular cluster XB span 1 -- 10 $\times 10^{37}$ erg s$^{-1}$, hence we expect high accretion rate PDS with no breaks. No breaks were  observed in the PDS of any of the globular clusters.

\begin{figure}[!b]
\centering
\resizebox{\hsize}{!}{\includegraphics[angle=270]{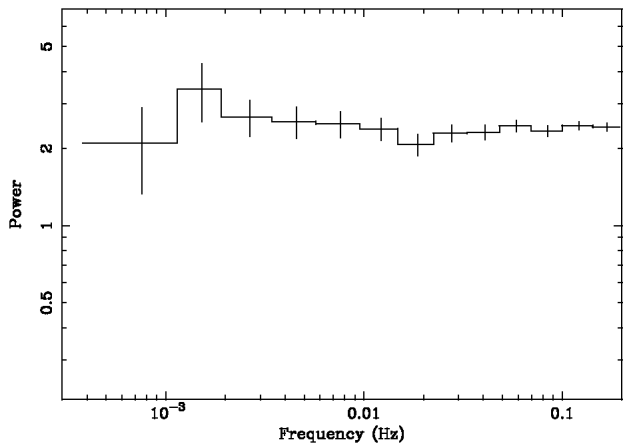}}
\caption{The noise-subtracted PDS of the combined lightcurves of the seven X-ray sources associated with globular clusters used to search for an artificial break in the PDS; no break is seen.} \label{gcpds}
\end{figure}

 Additionally, a lightcurve was produced by summing all the EPIC lightcurves of the globular cluster sources; the intensity of this summed lightcurve is $\sim$1.5 count s$^{-1}$, (c.f. 0.3--1 count s$^{-1}$ for RX\thinspace J0042.3+4115), hence the PDS had more than sufficient data to detect any instrumental break.  The PDS  of that lightcurve is presented in Fig.~\ref{gcpds}. A break in the PDS due to instrumental effects  would be present in all PDS and so adding the lightcurves of several sources should make the break more prominent; in fact, fitting a simple power law to the combined globular cluster PDS resulted in a good fit ($\chi^{2}$/dof of 0.6). A broken power law is not required to fit the globular cluster  data. We can thus be confident that the PDS breaks we have observed are inherent.

\section{Discussion}
The power density spectra observed from RX\thinspace J0042.3+4115 by XMM-Newton are flat with a break at $<$1 Hz, consistent with PDS from disc-accreting X-ray binaries XB at a low accretion rate, and inconsistent with PDS from high accretion rate XB \citep{vdk89,wv99}. In LMXB with neutron star primaries, the characteristic low accretion rate PDS appear in the island state, the lowest luminosity state \citep{vdk89}; an example is \object{1E\thinspace 1724$-$3045}, an X-ray burster that exhibited a low accretion rate PDS at a few 10$^{36}$ erg s$^{-1}$ \citep{ol98}. The fact that RX\thinspace J0042.3+4115 exhibits a low accretion rate PDS at luminosities exceeding 10$^{38}$ erg s$^{-1}$, and that none of the globular cluster sources (likely LMXB with neutron star primaries)  exhibit such PDS  at luminosities of 0.1 -- 1 10$^{38}$ erg s$^{-1}$ strongly suggests that the primary in  RX\thinspace J0042.3+4115 is unlikely to be a neutron star. We conclude that the primary of  RX\thinspace J0042.3+4115 is a black hole candidate.

Since the argument rests on the luminosity of  RX\thinspace J0042.3+4115, it is important to be sure of its location in M31. M31 is located well out of the plane of the Milky Way, in a sparsely populated region of sky, meaning that  RX\thinspace J0042.3+4115 must be within our Galaxy, or in M31 or a background AGN. We discount the possibility that  RX\thinspace J0042.3+4115 is an AGN and discuss possibilities where it is a  local binary below.  

 AGN are powered by disc accretion onto massive black holes, so we might expect them to exhibit similar PDS, scaled to  longer time-scales. \citet{utt02} examined the PDS of four AGN (\object{MGC-6-30-15}, \object{NGC\thinspace  5506}, \object{NGC\thinspace 3516} and \object{NGC\thinspace 5548}) and found that the PDS of the first three are analogous to low accretion rate XB, but with the break at a few 10$^{-6}$ Hz. These authors compare the PDS of these four AGN with that of the black hole XB Cygnus\thinspace X-1 and conclude that the break frequency is related to the mass. The break frequency of RX\thinspace J0042.3+4115 is several orders of magnitude higher than observed in these AGN and so we reject the possibility that it is a background AGN.
\begin{table*}[!t]
\centering
\caption{ A catalogue of point X-ray sources identified as globular clusters by \citet{K02} and used to test the hypothesis that the break in the PDS of RX\thinspace J0042.3+4115 is not instrumental. The best fit parameters of absorbed power law models to the spectra are given; the column headings have the same meanings as in Table ~\ref{specres}} \label{gc}
\begin{tabular}{llllllll}
\noalign{\smallskip}
\hline
\noalign{\smallskip}
Source & N$_{\rm H}$ & $\Gamma$ & n$_{\rm P}$ & $\chi^{2}$/d.o.f & F$^{0.3-10}$ & L$_{\rm 760}^{0.3-10}$ & Break?\\
\noalign{\smallskip}
\hline
\noalign{\smallskip}
\object{J\thinspace 004212.1+411758}& 0.35 & 1.66 & 7.5 10$^{-5}$ & 97/114 & 0.54 & 3.7& N\\
\object{J\thinspace 004218.6+411401} & 0.13 & 1.55 & 1.29 10$^{-4}$ & 306/301 & 1.0 & 7 & N \\
\object{J\thinspace 004226.0+411915} & 0.42 & 2.32 & 4.5 10$^{-5}$ & 56/73 & 0.22 & 1.5 & N\\
\object{J\thinspace 004232.1+411939} & 0.21 & 2.06 & 7.4 10$^{-5}$ & 153/127 & 0.4 & 2.8& N\\
\object{J\thinspace 004259.6+411919} & 0.13 & 1.98 & 1.4 10$^{-4}$ & 275/250 & 0.78 & 5.4 & N\\
\object{J\thinspace 004310.6+411451} & 0.10 & 1.74 & 2.12 10$^{-4}$ & 430/429 & 1.4 & 9.8 & N\\
\object{J\thinspace 004337.2+411443} & 0.14 & 1.70 & 1.34 10$^{-4}$ & 268/241 & 0.93 & 6.4& N\\
\noalign{\smallskip}
\hline
\noalign{\smallskip}
\end{tabular}
\end{table*}

If  RX\thinspace J0042.3+4115 is actually a foreground object, then it probably belongs to the disc population and is located within $\sim$ 1 kpc. Hence the X-ray luminosity would be $\sim$6 orders of magnitude fainter, at $\sim10^{32}$ erg s$^{-1}$, or fainter if closer than 1 kpc. We know from the PDS that RX\thinspace J0042.3+4115 is powered by disc accretion, but this X-ray luminosity is several orders of magnitude too low for a persistent LMXB \citep{lvv95}. Hence RX\thinspace J0042.3+4115 would have to be a cataclysmic variable (i.e. with a white dwarf primary) or a soft X-ray transient in quiescence.

All known Galactic cataclysmic variables have optical/UV counterparts \citep[e.g][]{down01} and while RX\thinspace J0042.3+4115 was associated with an optical source using the Einstein position \citep{tf91}, the improved resolution of Chandra placed RX\thinspace J0042.3+4115 away from any optical counterpart \citep{K02}. The nearest object in the  \citet{hai94} optical catalogue is at 00$^{\rm h}$~42$^{\rm m}$~ 28$^{\rm s}$~+41$\degr$~15$\arcmin$~35$\farcs$7, i.e. just 5$^{\rm s}$ away and has  an apparent visual magnitude of m$_{\rm V}$ = 20.075. Hence an optical counterpart to RX\thinspace J0042.3+4115 must  be fainter than m$_{\rm V}$ = 20. A pessimistic estimate of the  absolute visual magnitude for a non-magnetic CV is M$_{\rm V}$ = 9, considering only disc emission  and an accretion rate of 10$^{-11}$ M$_{\odot}$ yr \citep{war95}; hence at 1 kpc, the apparent magnitude would be brighter than 19 and would have been seen in the \citet{hai94} survey. Hence we can rule out the possibility that RX\thinspace J0042.3+4115 is a normal non-magnetic CV. The observed spectra  from RX\thinspace J00422.3+4115 are consistent with Galactic magnetic CVs, but the X-ray luminosity at 1 kpc would be two orders of magnitude lower than normal. The system would have to be one of the theoretically implied, post period minimum  CVs with a brown dwarf secondary \citep{kb99}.

If RX\thinspace J00422.3+4115 is a local soft X-ray transient (SXT) in quiescence, its primary could be a neutron star or a black hole. SXTs typically brighten in X-rays  by a factor of up to 10$^{7}$ over a week, then decay into quiescence over the course of a year; successive outbursts are  usually separated by years to decades \citep[see review in e.g.][]{ts96}. Observed spectra from SXTs in quiescence are often modelled using advection dominated accretion flow (ADAF) in the inner disc \citep[e.g.][]{ngm97,sut02}. Since ADAFs are radiatively inefficient, a neutron star primary will radiate energy donated by accreting material impacting on the surface, while a black hole primary will simply swallow the accreting material; indeed quiescent neutron star SXTs have been observed to be 2 orders of magnitude brighter than quiescent black hole SXTs \citep[][ and references within]{sut02}. Comparing the implied luminosity of  10$^{32}$ erg s$^{-1}$ for RX\thinspace J00422.3+4115 with the known minimum luminosities of black hole and neutron star SXTs \citep{ham02}, we find that if RX\thinspace J00422.3+4115 was a local SXT, it would be likely to have a black hole primary. 

Accepting that the RX\thinspace J00422.3+4115 is a black hole binary in M31, we must account for its being in the low accretion rate state at a luminosity as high as a few 10$^{38}$ erg s$^{-1}$. The mostly likely explanation is that the low accretion rate state is limited to less than a certain fraction of $\dot{M}_{\rm Edd}$,  the mass accretion rate corresponding to the Eddington limit, rather than $\dot{M}$ itself; hence RX\thinspace 00422.3+4115 would only need to be more massive than a neutron star to allow a low accretion rate state at a higher accretion rate (observed luminosity).  GS\thinspace 1124-683 (Nova Muscae 1991) exhibited a low-accretion rate PDS at a 1.2--37 keV luminosity of $\sim 10^{38}$ erg s$^{-1}$, for a distance of 4 kpc \citep{miy93,mg99}; the mass of the primary in this system has been determined to be 7.0$\pm$0.6 M$_{\odot}$ \citep{gel01}, so in this case $\dot{M}$ $\sim$0.1 $\dot{M}_{\rm Edd}$. Hence, a scenario in which RX\thinspace J00422.3+4115 has a massive ($\sim$10 M$_{\odot}$) primary accreting at $\sim$0.1 $\dot{M}_{\rm Edd}$ is feasible and entirely consistent with the  observed properties.
 
If the break frequency is related to the mass of the primary as supposed by \citet{utt02}, the similar break frequencies of Cygnus X-1 [0.04--0.4 Hz \citep{bh90}] and RX\thinspace J0042.3+4115 (0.05--0.09 Hz) suggests that the mass of the black hole in Cygnus X-1 [10 M$_{\odot}$ \citep[][]{her95}] is similar to that in RX\thinspace J0042.3+115,  supporting the idea of a $\sim$10 M$_{\odot}$ black hole primary in  RX\thinspace J0042.3+4115.
 
To conclude, the three scenarios discussed for RX\thinspace J0042.3+4115 are: that RX\thinspace J0042.3+4115 is a post period minimum CV with a brown dwarf secondary; that RX\thinspace J0042.3+4115 is a luminous black hole binary in M31; that RX\thinspace J0042.3+4115 is a local black hole SXT with intrinsic absorption imitating the absorption found in M31. We consider that it is most likely to be placed in M31. In this case, the system could be a microquasar, i.e. with a low mass companion; it would then be an analogue of the Galactic microquasar GRS\thinspace 1915+105.

\begin{acknowledgements}
 The authors would like to thank Carole Haswell for useful conversations and Andrew Norton for the code used to model the PDS. This work is supported by PPARC.
\end{acknowledgements}
\bibliographystyle{aa}
\bibliography{m31}
\end{document}